# Large and Flat Graphene Flakes Produced by Epoxy Bonding and Reverse Exfoliation of Highly Oriented Pyrolitic Graphite

Vincent Huc[1], Nedjma Bendiab[2,*], Noël Rosman[3], Thomas Ebbesen[4], Cécile Delacour[5] and Vincent Bouchiat[5,+]

1-ICMMO-CNRS, Université Paris-Sud, 91405 Orsay, France
2-Laboratoire de Spectrométrie Physique, Université Joseph Fourier, 38402 Saint Martin d'Hères, France.
3-LEPMI, ENSEEG, INPG, F38402 Saint Martin d'Hères, France.
4-ISIS, Université Louis Pasteur, 8, Allée Gaspard Monge BP 70028, 67083 Strasbourg Cedex, France.
5-Institut Néel, CNRS-Grenoble, 25 av. des Martyrs, 38042 Grenoble Cedex, France.

* Now at Institut Néel, CNRS-Grenoble
+ On sabbatical leave at Department of Physics, University of California, CA-91720 Berkeley, USA.

### Abstract

We present a fabrication method producing large and flat graphene flakes that have a few layers down to a single layer based on substrate bonding of a thick sample of highly oriented pyrolytic graphite (HOPG), followed by its controlled exfoliation down to the few to single graphene atomic layers. As the graphite underlayer is intimately bonded to the substrate during the exfoliation process, the obtained graphene flakes are remarkably large and flat and present very few folds and pleats. The high occurrence of single layered graphene sheets having tens of micron wide in lateral dimensions is assessed by complementary probes including spatially resolved Micro-Raman Spectroscopy, Atomic Force Microscopy and Electrostatic Force Microscopy. This versatile method opens the way of deposition of graphene on any substrates including flexible ones.

## 1. Introduction

Many methods have been developed in the last decades to obtain few layers of graphene. They are based on very different physicochemical processes and can sorted in different categories : Dry cleaving techniques of highly oriented pyrolytic graphite (HOPG) or graphite single crystals (either obtained by peeling off bulk graphite using adhesive tape [1] or cleavage by mechanical rubbing [2]), electrostatic deposition of graphene [3], chemical [4] or thermal [5] exfoliation , chemical decomposition[6] of graphitic materials, thermal decomposition and reconstruction on silicon carbide [7], [8], [9] or metal surfaces [10] in Ultra-High Vacuum or chemical vapour deposition [11]. However this rapidly expanding field [12] was triggered by the discovery of a very simple fabrication process based on the deposition of mesoscopic graphite sample created by the exfoliation of HOPG sample followed by a random deposition [1]. The success of that method is mainly due to the easy characterization of the number of layers using optical microscopy [13, 14]. However recent scanning probe studies [15] shows that the graphene layers are far from being flat at the atomic level. Graphene deposited on silica is indeed known to give rise to a corrugated surface [15, 16] and the influence of the charged silica underlayer on the graphene electronic transport properties is today recognized as a critical parameter that limits [17] the electron mobility within the graphene layer. Indeed Coulomb scattering induced by the trapped charge on Silica is the major phenomenon that limits the electron transport in graphene for a Fermi level positioned near the Dirac point. New fabrication methods of few layers graphene on substrates on with an expected density of charge traps lower than in silica are therefore highly interesting for studying quantum transport properties.

Several techniques including transfer printing [18, 19] of graphene has been elaborated to produce graphene sheets on surface other than silica, but does not address simultaneously the flatness and surface area issue. We propose in this pqper a method based on substrate bonding. Production of thin crystalline multilayers by bonding [20] is indeed an active and mature field of research that has already lead to important industrial applications [21],[22],[23]. This paper is inspired from that field, namely to produce flat graphene layers firmly bonded to an insulating underlayer during the exfoliation process.

## 2. Preparation Technique

The proposed method (Fig. 1) is somewhat reminiscent of the process developed for the industrial production of thin silicon-on-insulator wafers [22]. A thin "layered" crystalline film is bonded and firmly mechanically supported during the thinning. Our process mainly consists of two steps (fig. 1): Firstly, graphite bonding is realized using a thin layer of epoxy resist which is subsequently cured under pressure. Secondly the bonded graphite is exfoliated using the scotch tape technique. A freshly cleaved HOPG crystal is glued using very thin epoxy-based glue (see Suppl. Info. for details). After curing, the glue leaves creates a transparent, void and bubble-free interlayer between the substrate and the HOPG sample (see SEM picture in fig S2 in Suppl. Info. and fig. 2). This fabrication method can be refined by using a pre-cut (<100µm thick) flexible HOPG sample as the top layer. This meteod may appear as a refinement of the exfoliation process known as the "scotch-tape" technique [1]. However significant differences do exist with respect to that well-known method. Firsly one should note that the useful graphene sheets is the one left on the (fist) adhesive side (i.e. the epoxy underlayer) after the curing step (hence the name "reverse exfoliation"). This leads ton larger and flatter flakes with very few pleats (as seen in Fig.2). The second one is that a uniaxial compression is applied to the sample during the adhesive curing step, in order to increase the flatness of the adhesive interlayer. The third one is that the adhesive used in this study was found to be compatible with conventional micro/nanofabrication techniques including lithographic and etching processes. This allows transport measurements to be performed in-situ directly on the bonded monolayer.

## 3. Characterization of the graphene samples

During our first trials, the epoxy thickness was not controlled but remained in the sub-10µm range. This lack of control in the sub-micrometer range prevents us from using the usual optical interferometry method [13], [14] as a tool for optical determination of the number of layers. Therefore, in order to assess the quality and number of layers of the graphene produced by our technique, we used local characterization techniques (near-field microscopies and Raman spectroscopy), both of them being sufficiently reliable to obtain graphene fingerprints.

## 3.a Micro-Raman spectroscopy

The most efficient way to characterize the samples produced by our method was found to be micro-Raman spectroscopy. A detailed spectrum can be acquired at registered positions with micrometer scale accuracy. Raman spectra of graphene have typical signatures [24],[25]. Its in-depth analysis can allow the precise determination of the number of layers.

The Raman spectrum of graphite is composed of a strong band at 1582 cm$^{-1}$, which has been assigned to the in-plane $E_{2g}$ zone centre mode (G band). We also observed an additional first order line at ~1352 cm[-1] (D band) which have been assigned to non Brillouin zone-center phonon [26]. This D band is induced by disorder and is highly dispersive [26, 27]. Finally, the associated overtone 2D band around 2700 cm[-1] is pronounced even in the absence of the D band. Hence, vibrational and electronic properties of graphite are dominated by the $sp^2$ nature of the strong intra-plane covalent bonds. Raman spectra for various numbers of layers can be compared quantitatively and qualitatively. As described recently [24, 25] , we can extract the number of graphene layers by precisely measuring the G band position.

Figure 3 shows two high-frequency Raman spectra associated with two microscope images. These spectra were recorded successively, by moving the x-y stage to the next position. Raman spectra of the bare epoxy bonding layer have been acquired (data not shown) and show no peaks in the measured wavelength range. The spectrum on Fig3a (bottom curve, black one), is typical of a thick graphite sample with a large number of layers..In comparison, the spectrum on fig 3, (top curve, red one) shows a G band up shifted to 1587 cm[-1]. This latter is a signature of a monolayer graphene[23, 24]. In addition, the 2D mode measured for n=∞ and for n=1 (Fig. 3c) confirms the strong dependence of the 2D mode with n according to a double resonance model. As expected for one monolayer, the 2D mode is fitted with only one peak with a FWMH $\Gamma$= 22 cm[-1] whereas for graphite, two large bands are required. It is worth mentioning that the amount of single/few layer(s) graphene on the surface was found to be much larger than the amount of multilayered graphite.

The bottom spectrum on Fig. 3a  (in black) is typical of a graphite spectrum having a large number n of graphene layers . The bottom spectrum  on Fig. 3c acquired on the same samples shows a G band up shifted to 1587 cm$^{-1}$. In comparison, This spectrum is a signature of a monolayer graphene. In addition, the 2D mode measured for n=∞ and for n=1 (Fig. 3) confirms the strong dependence of the 2D mode with n according to a double resonance model. As expected for one monolayer, the 2D mode is fitted with only one peak with a FWMH $\Gamma$ = 22 cm$^{-1}$ whereas for graphite two large bands are used. It is worth mentioning that the amount of single/few layer(s) graphene on the surface was found to be much larger than the amount of multilayered graphite. Along with the large observed size of these single graphene domains of tens of microns size, this constitutes a significant added value compared with others processes for the production of graphene.

### 3.b Scanning probe microscopy study

Atomic Force Microscopy (AFM) images were recorded in sample areas previously identified as single layer graphene by Micro-Raman. A AFM picture of the acquired in the area previously analyzed by Raman in Fig. 3 is shown in Fig. 4. Localization of the graphene sheets was not straightforward, as the graphene sheets are moulded into the resist during curing and their removal during the exfoliation step leaves some "ghost" footprints with a surface topology very similar to the real graphene sheets. Discrimination between real graphene and molded epoxy surfaces could however be obtained using a voltage biased conductive AFM tip. In this so-called mode of Electrostatic Force Microscopy (EFM), the transition between the insulating resist and the conducting graphene can be easily revealed since the graphene sheet provide a surface capacitance that shields the local electric field produced by the biased tip. Fig. 5 shows simultaneously acquired topographic and EFM images on an edge between bonded graphene and bare epoxy. The EFM image shows a highly contrasted step at the graphene/epoxy transition. The measured step height 0.4 is in accordance with the expected one for a single graphene layer (0.35nm). Similar charging contrast can be obtained using SEM imaging (fig. 2).

The graphene surface is relatively smooth (fig.4), in sharp contrast with the folds and pleats commonly observed with solution-processed graphene. Only few cracks are observed, thus confirming the usefulness of mechanically supporting the graphene single layer to accommodate for the high mechanical constraints associated with the exfoliation process.

Some residues presumably coming from the scotch tape glue are however observed (white dots on figure 4b). Such residues are generally associated with the thinnest layers, when the adhesive layer of the scotch tape is in contact with the graphene directly bonded to the surface. Such a phenomenon seems to be related to the competition between the substrate/graphene and tape/graphene adhesive forces during the exfoliation. Roughness analysis performed on the single layer area (according to the Raman signal) gives a RMS roughness of 0.16 nm. This value is lower by a factor of two than typical results obtained on graphene deposited on silica with the conventional technique [15]. Electrostatic Force Microscopy was also used to evidence the presence of single graphene sheets on the surface (Figure 5. Left: topographic image; right EFM image). The very different electrical behaviours between the insulating substrate on one hand and the metallic graphene on the other hand results in a clear potential contrast between these two surfaces.

SEM analysis of the same sample (fig. 2) confirms the conclusions drawn from the AFM study. The graphene domains are rather smooth, with relatively few pleats and folds.

the presented method should allow the reproducible fabrication of graphene layers on flexible surfaces. The thickness of the adhesive layer was found to be about in the micron range. The reduction of this thickness by resist dilution seems feasible.

**3c Deposition of Palladium electrodes**

Different adhesion promoters were tested during this work. The M-Bond 610 type adhesive was found to be compatible with the deposition of palladium electrodes by conventional lithographic procedure (see suppl. info). Ohmic contacts were reproducibly obtained, with kilo-ohms contact resistances

**4. Conclusion**

In conclusion, a new process for the formation of graphene single layers is described. This process allows for the deposition of large and flat graphene single layers. The characterization of the samples shows the high occurrence of single layer graphene. The relatively large extension of single layer graphene sheets is assessed by detailed scanning probe microscopy, together with the observation of both the well characterized G-Band Frequency shift and the concomitant evolution of the D band in the Raman spectra.

Finally this simple technique appears as rather versatile and should allow for the transfer of graphene (possibly of different origins, i.e. Silicon carbide or Nickel supported) on a vast variety of samples including transparent and/or flexible ones, while keeping a flat surface. This is for example an interesting requirement for the development of graphene based conductive films for flexible electronics including transparent electrodes [28, 29] and nanostructured carbon-based sensors [30]

In conclusion, we have presented a simple technique of fabrication of large graphene flakes. An interesting feature of that technique compared to the classical scotch tape technique is that here the process presents a self-limiting character: Indeed the last layer of graphene is strongly bonded on the epoxy, and additional application of the tape is mostly inefficient to remove it. This explains the occurrence of large single layers of graphene obtained by this technique.

**Acknowledgments**

This work has been partly supported by D.G.A. and Région Rhône-Alpes. V.B. acknowledges support from the Miller Institute of Basic Science.

The Authors would like to thank Emmanuel Andre for help with the epoxy based adhesives, Caglar Girit, Irina Ionica and Wolfgang Wernsdorfer for stimulating discussions.

# **Figures**

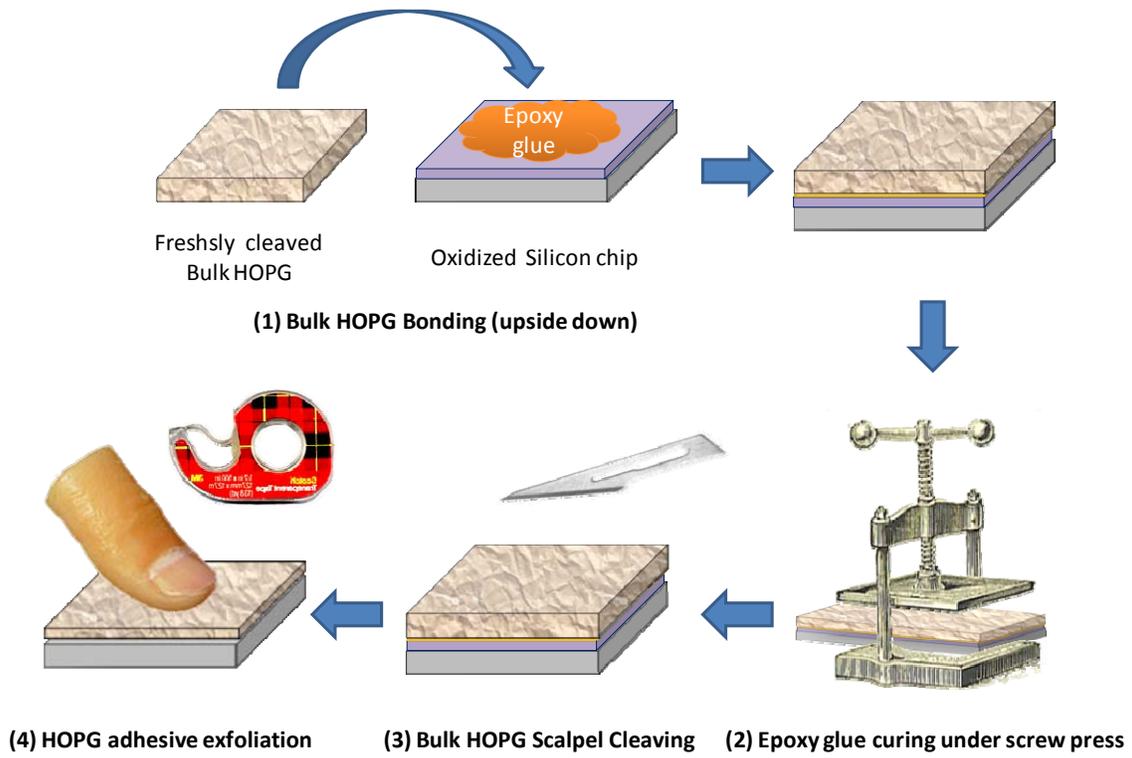

Figure 1: Schematics of the process flow depicting the reverse exfoliation of graphite leading to graphene layers.

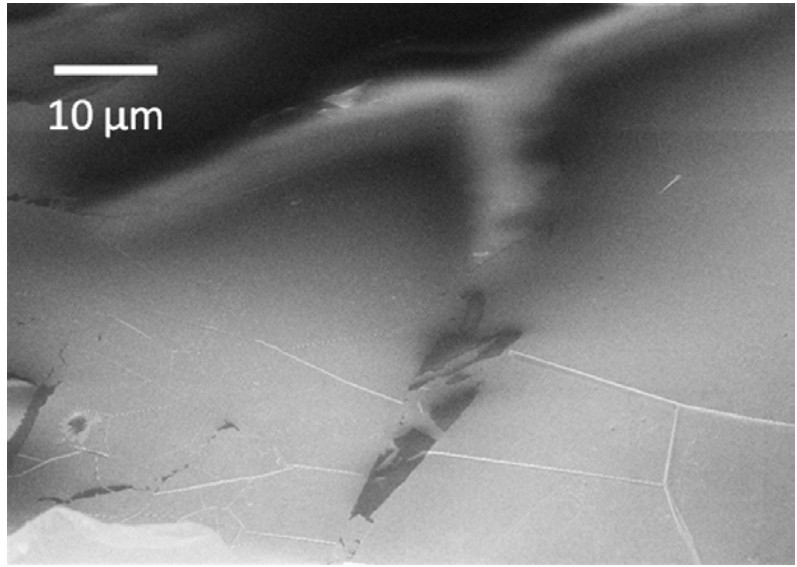

Figure 2: SEM micrograph of a typical few-layers graphene sample obtained by the reverse exfoliation method. White areas in the top of the image originate from charging in the SEM of the uncovered epoxy resist. Note the large extension (>100µm²) of defect free areas.

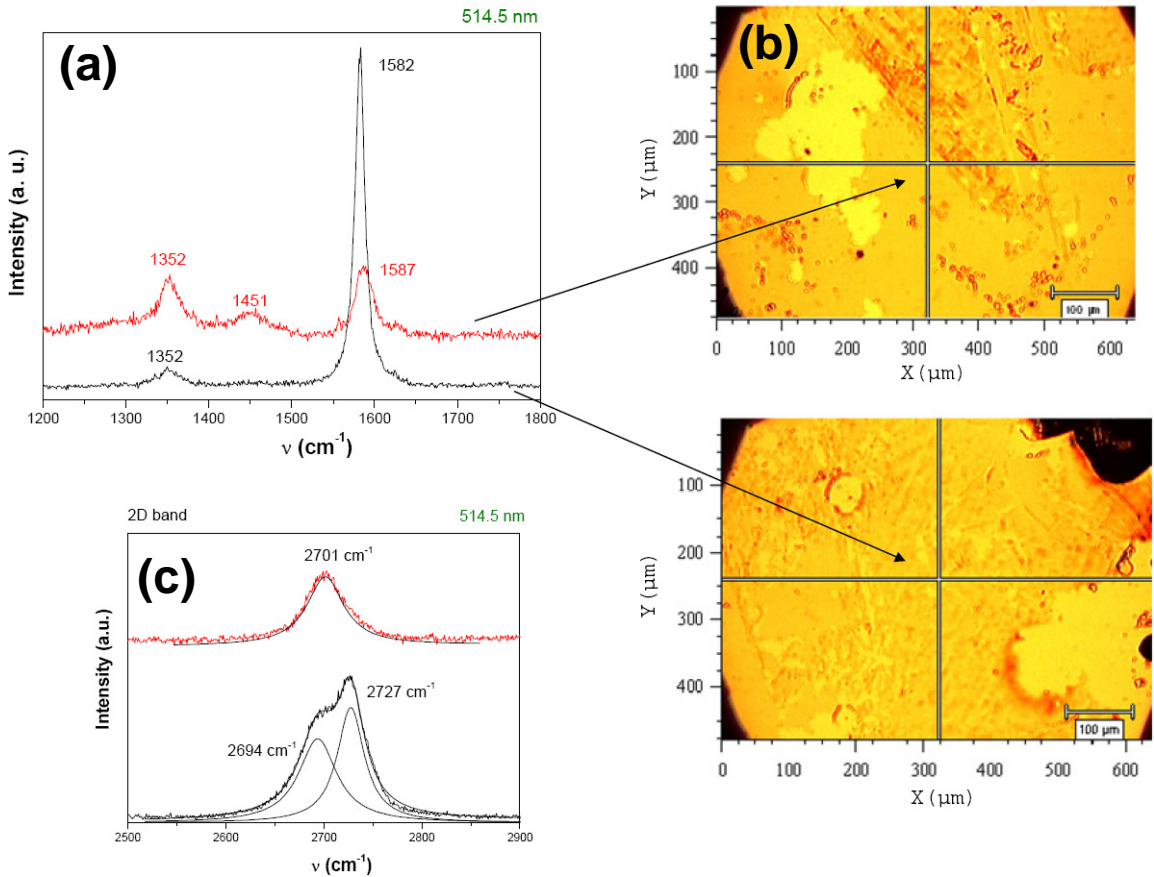

Figure 3: Raman spectra of the graphene layers obtained by reverse exfoliation. (a) Compared Raman spectra for a single layer graphene (red curves) and for a thick HOPG Flake (graphite, black curves) acquired at different areas (shown as arrows on the right image) on the same sample. (b) Optical micrographs of the tested sample. The probed area is at the centre of the reticule for each spectrum. (c) Raman spectra of the 2D mode on the same samples.

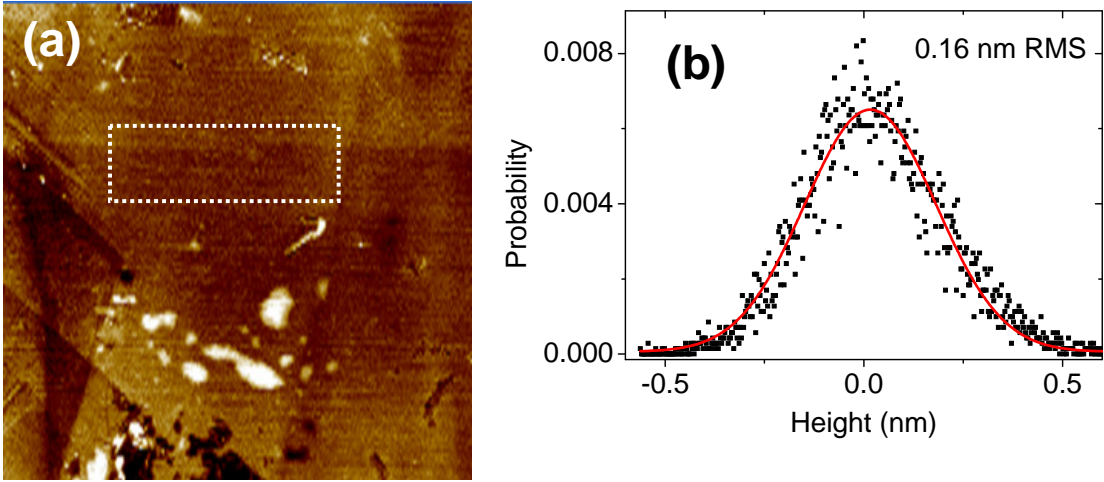

Figure 4: (a) Atomic Force Micrograph (2μm scan) of a reverse exfoliated graphene sample taken at the boundary with a bare epoxy surface. Scan is performed at the position where the Raman signal indicate a single layer thickness. Height colour scale is 5nm. The dotted with rectangle in left depicts the area where roughness analysis is performed. (b): Height distribution of the single layer graphene film on epoxy (data taken from inside the dotted rectangle). The RMS roughness extracted from the Gaussian fit is 0.16nm.

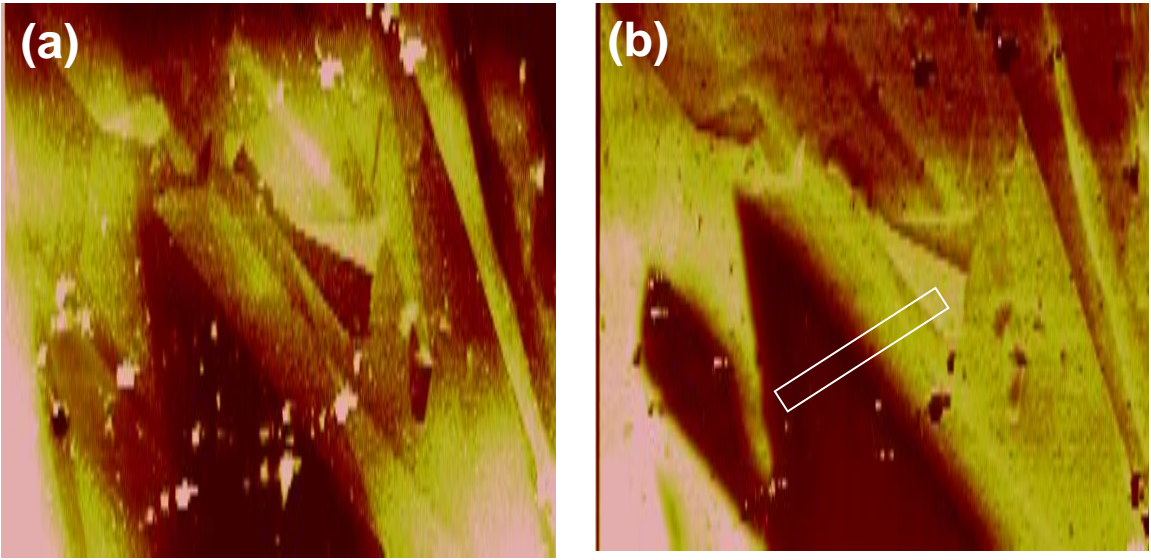

Figure 5. Comparison of a exfoliated graphene surface of a few atomic layer thickness near the boundary of a bare epoxy layer with Atomic Force (a) and Electrostatic Force (b) Microscopies (topographic image vertical scale 5nm, EFM image Lift height = 5 nm, Tip bias = 2V, 6 μm scan). The white rectangle corresponds to a single atomic layer step.

# References

[1] Novoselov K S, Geim A.K., Morozov S.V., Jiang D, Zhang Y, Dubonos S V, I.V. G and Firsov A.A. 2004 *Science* **306** 666

[2] Novoselov K S, Jiang D, Schedin F, Booth T J, Khotkevich V V, Morozov S V and Geim A K 2005 *Proceedings of the National Academy of Sciences* **102** 10451-3

[3] Sidorov A N, Yazdanpanah M M, Jalilian R, JOuseph P, WCohn R and Sumanasekera G U 2007 *Nanotechnology* **18** 135301

[4] Dresselhaus M S and Dresselhaus G 2002 *Adv. Phys.* **51** 1–186

[5] Hannes C. Schniepp, Li J-L, McAllister M J, Sai H, Herrera-Alonso M, Douglas H. Adamson, Prud'homme R K, Car R, Saville D A and Aksay I A 2006 *J. Phys. Chem. B* **110** 8535–9.

[6] Dujardin E, Ebbesen T W, Krishnan A and Treacy M M 1998 *AdV. Mater.* **10** 1472

[7] Bommel A J V, Crombeen J E and Tooren A V 1975 *Surface Science* **48** 463

[8] Forbeaux I, Themlin J M and Debever J M 1998 *Physical Review B* **58** 16396

[9] Berger C, Song Z, Li T, Li X, Ogbazghi A, Feng R, Dai Z, Marchenkov AN, Conrad E, PN F and de Heer W 2004 *J. Phys. Chem. B* **108** 19912

[10] McConville C F, Woodruff D P and Kevan S D 1986 *Surface Science* **171** L447

[11] Wang J J, Zhu M Y, Outlaw R A, Zhao X, Manos D M, Holloway B C and Mammana V P 2004 *Appl. Phys. Lett.* **85** 1265

[12] Geim A K and Novoselov K S 2007 *Nature Materials* **6** 183

[13] Blake P, Hill E W, Neto A H C, Novoselov K S, Jiang D, Yang R, Booth T J and Geim A K 2007 *Appl. Phys. Lett.* **91** 063124

[14] Abergel D S L, Russel A and Falko V I 2007 *Appl. Phys. Lett.* **91** 063125

[15] Ishigami M, Chen J H, Cullen W G, Fuhrer M S and Williams E D 2007 *Nano letters* **7** 1643

[16] Meyer J C e a 2007 *cond-mat/0703033*

[17] Du X, Skachko I, Barker A and Andrei E Y 2008 *condmat arXiv:0802.2933v1*

[18] Chen J-H, Ishigami M, Jang C, Hines D R, Fuhrer M S and Williams E D 2007 *Advanced Materials* **19** 3623-7

[19] Ritter K A and Lyding J W 2008 *Nanotechnology* **19** 015704

[20] Niklaus F, Stemme G, Lu J-Q and Gutmann R J 2006 *J. Appl. Phys.* **99** 031101

[21] M. Bruel et al. 1995 *Electronics Letters* **31** 1201

[22] Bruel M *US Patent 5374564*

[23] http://www.soitec.com

[24] Gupta A, Chen G, Joshi P, Tadigadapa S and Eklund P C 2006 *Nano letters* **6** 2667

[25] Ferrari A C, Meyer J, Scardaci C, Casiraghi C, Lazzeri M and Mauri F 2007 *Phys. Rev. Lett.* **97** 187401

[26] Thomsen C and Reich S 2000 *Phys. Rev. Lett.* **85** 5214

[27] Maultzsch, Reich and Thomsen 2004 *Phys. Rev. B* **70** 155403

[28] Wang X, Zhi L and Mullen K 2008 *Nano Lett.* **8** 323-7

[29] Blake P, Brimicombe P D, Nair R R, Booth T J, Jiang D, Schedin F, Ponomarenko L A, Morozov S V, Gleeson H F, Hill E W, Geim A K and Novoselov K S *arxiv:0803.3031*

[30] Bradley K, Gabriel J-C and Gruner G 2003 *Nano Lett.* **3** 1353